# Anisotropic re-entrant spin-glass features in a metallic kagome lattice, $Tb_3Ru_4Al_{12}$

E. V. Sampathkumaran,[a,*] Kartik K Iyer,[a] Sanjay K Upadhyay,[a] and A.V. Andreev[b]

[a]*Tata Institute of Fundamental Research, Homi Bhabha Road, Colaba, Mumbai 400005, India*
[b]*Institute of Physics, Academy of Sciences of the Czech Republic, Na Slovance 2, 182 21 Prague, Czech Republic*

**Abstract**
We present the results of ac and dc magnetic susceptibility and isothermal magnetization measurements ($T$= 2-300 K) on the single crystals of a metallic kagome lattice, $Tb_3Ru_4Al_{12}$, reported recently to undergo reentrant magnetism with the onset of long range antiferromagnetic order below ($T_N$ = 22 K). The magnetization data obtained on the crystal with the $c$-axis orientation along magnetic-field reveal spin-glass-like characteristics near 17 K (below $T_N$). However, for the orientation along basal plane, such glassy anomalies are not observable above 2 K. In this respect, this compound behaves like an 'anisotropic' re-entrant spin-glass. Possible implications of this finding to the field of 'geometrically frustrated magnetism' are considered.

*Corresponding author: sampathev@gmail.com

1. **Introduction**

'Frustrated magnetism' [1] arising from a tendency to hinder simple macroscopic long-range ordering by the geometrical arrangement of magnetic ions in the lattice is one of the important topics of research in condensed matter physics. Apart from the reduction of magnetic ordering temperature with respect to paramagnetic Curie temperatures, consequences of complete suppression of magnetic ordering leading to other novel ground states like spin-ice (in pyrochlores) and spin-liquid [2] have been demonstrated in the literature. Many geometrically frustrated systems have been found to exhibit spin-glass freezing as well [3]. These include triangular lattices, tetrahedral compounds like pyrochlores, and kagome lattices. During early stages of the development of the field of 'geometrically frustrated magnetism', there were 'fully frustrated models' that is, not invoking disorder, attributing spin-glass anomalies to geometrical frustration [4, 5]. However, many theoretical works [3, 6-10] subsequently emphasized the need to invoke crystallographic disorder, even at low levels, to explain experimentally observed spin-glass freezing in many systems. Judged by the literature, a vast majority of researchers seem to agree with this.

However, there are still some models which bring out intrinsic nature of spin-glass anomalies (that is, without disorder). Such alternative approaches also seem to be reasonable, judged by the fact that 'structural glasses' arising out of geometrical frustration in supercooled liquids do not actually require crystallographic disorder [11]. Cepas and Canals proposed a simple spin model in which a 'dynamical spin-glass freezing' in the absence of quenched disorder [12] is predicted. In this model, spin relaxation develops spontaneously two time scales below a crossover temperature, with fast-moving and slow-moving regions, which may look like frozen for certain experimental time scales. Chandra et al in fact long ago theorized [5] that anisotropic kagome antiferromagnets are not conventional antiferromagnets and such a lattice could be a natural setting for glassiness without the role of disorder. Kagome lattices are generally characterized by magnetic ions placed at the corners of a hexagon and each face of the hexagon forms a triangle with another magnetic ion. While a larger number of well-studied kagome compounds, investigated for such magnetic frustration phenomena, are known to be insulators, this structure, with nearest neighbor antiferromagnetic interaction to favor magnetic frustration, is not common in a metallic environment, barring a few, e.g., those crystallizing in ZrNiAl structure [13] and $GdInCu_4$ [14]. In metallic systems, Rudermann-Kittel-Kasuya-Yosida indirect exchange interaction is operative, which is absent in the insulators. We therefore consider it worthwhile to search for kagome lattices in metallic environment with anisotropic spin-glass anomalies, which would add to the data base to eventually test the predictions of Refs. 5 and 12.

In this respect, the compounds of the type, $R_3Ru_4Al_{12}$ (R= Rare-earths) [15-21], crystallizing in a hexagonal structure (space group, $P6_3/mmc$), containing $R$ ions in the kagome network are of importance. Recently, detailed magnetic investigations on the polycrystals of $Tb_3Ru_4Al_{12}$ have been reported, which, apart from confirming long-range antiferromagnetic ordering at 22 K [17], revealed re-entrance to a glassy-like phase around 17 K. This prompted us to reinvestigate the single crystals of this compound by *ac* and *dc* magnetization (*M*) measurements. We report here that the characteristic features of glassy features could be observed in the magnetization data below about 17 K, however, only for the orientation of the *c*-axis of the crystal along the direction of the magnetic-field (*H*). Thus, this work identifies a kagome lattice with anisotropic spin-glass features in the bulk measurements, possibly making the physics of this compound relevant to the theory in Ref. 5.

2. **Experimental details**

Single crystal of the title compound used for ac and dc magnetization studies is the same as those employed in Ref. 17. Its quality and orientation were checked by back-scattered Laue pattern (Fig. 1). Temperature (*T*) dependencies of dc $\chi$ (2 – 300 K) in 100 Oe and 5 kOe and ac susceptibility ($\chi$) for two different orientations of the crystal, $H//c$ and $H\perp c$, were obtained with the help of a commercial (Quantum Design) superconducting quantum interference device. In addition, isothermal remnant magnetization ($M_{IRM}$) measurements on the single crystal were performed with the same magnetometer at selected temperatures. In the case of ac $\chi$ measurements, the ac field was 1 Oe and the frequencies ($\nu$) employed were 1.3, 13, 133 and 1333 Hz.

3. **Results**
**3.1. Dc magnetization**

The temperature dependencies of dc $\chi$, measured in 5 kOe for the zero-field-cooled (ZFC) condition and in 100 Oe for ZFC and field-cooled (FC) conditions, for the two orientations of the crystal are shown in Fig. 2. Zero-field-cooling was carried out from 100 K. It is transparent from the figures that the values of $\chi$ for the basal plane orientation are at least an order of magnitude smaller compared to those for $H//c$. This is apart from the qualitative differences in the shapes of the curves for these orientations. This establishes strong anisotropy of magnetism of this compound, as suggested in Ref. 17 from the data collected in much high fields. In support of this, we have also measured isothermal $M$ at 2 K up to 50 kOe, which are shown in Fig. 3(a) and 3(b). The results are in good agreement with those reported in Ref. 17; for example, there is a hysteretic metamagnetic transition near 10 kOe for $H//c$. This metamagnetism is absent for $H\perp c$ in the field-range of measurement, and the values of $M$ at high fields (say, at 50 kOe) are significantly low, possibly due to crystal-field effects. Inverse $\chi$ (Figs. 2(a) and 2(b)) measured in a field of 5 kOe, is linear above 125 K, and a Curie-Weiss fit yields a larger effective moment of ~10.77 and ~10.04 ± 0.05 $\mu_B$/Tb with respect to the free ion value of $Tb^{3+}$ (9.72 $\mu_B$) for $H//c$ and $H\perp c$ respectively. This discrepancy can be attributed to a contribution from the polarization of the conduction electrons by the large Tb moment. It is possible that Ru 4d band acquires a moment, exhibiting itinerant magnetism when Tb is magnetically ordered. At low temperatures, there is a sharp upturn of $\chi$ below 22 K for $H//c$ as measured in 100 Oe (Fig. 2c) due to the onset of long-range magnetic order, though the sharpness is smeared in the curve obtained in 5 kOe (Fig. 2(a)). However, a broad peak only appears at a higher temperature (around 30 K) for the perpendicular orientation (Fig. 2(d)), without any upturn at 22 K. There is a weak upturn at a lower temperature range (<15 K), indicating complexity of magnetism. We believe that this weak upturn cannot be due to any impurity, as otherwise the net magnitude of this low temperature upturn for both the directions (say, in 5 kOe) should be similar. These results establish anisotropic nature of the magnetic ordering with the $c$-axis shown to be the easy axis of magnetic alignment.

In the case of $H//c$, the spin-glass feature seen in the polycrystals [21] is reproducible in the sense that ZFC and FC curves, obtained in 100 Oe, tend to deviate near 17 K, with the peak appearing in ZFC curve only at 14.7 K. Though ZFC-FC curves also marginally deviate from each other below 12 K for $H\perp c$, the absence of a peak in ZFC curve around 15 K supports anisotropy in possible glassiness. From the above data, it is inferred that the characteristic temperature associated with the onset of (anisotropic) glassiness is close to 15 - 17 K. Finally, we would like to mention that the hysteresis of FCC and field-cooled-warming curves observed in polycrystals (Fig. 3(b) in Ref. 21) could be confirmed by the measurements on single crystals as well (not shown here).

### 3.2. Isothermal Remnant Magnetization

We have measured $M_{IRM}$ at selected temperatures below $T_N$= 22 K to look for features characterizing spin glasses. For this purpose, the crystal was cooled in the absence of an external field to the desired temperature, switched on a field of 5 kOe for 5 mins and then the $M_{IRM}$ data was collected as a function of time ($t$). The time decay of $M_{IRM}$ is plotted in Fig. 3(c) and 3(d). The $M_{IRM}$ values at $t$= 0 (that is, the time at which $H$ reaches zero) for temperatures well below 17 K are rather large for $H//c$, compared to those for $H\perp c$, at a given temperature. The $t$=0 $M_{IRM}$ values are typically 0.0186, 0.002 and 0.000383 emu/g at 1.8, 8, and 15 K respectively for $H//c$. The values for $H\perp c$ at 1.8 and 3 K are 0.00696 and 0.00333 emu/g respectively. The point of emphasis is that, for $H//c$, the $M_{IRM}$ undergoes a slow decay at 1.8 and 8 K, reducing by about 30% after about 3 h. This decay is absent at higher temperatures, say, at 15 K and the value of $t$= 0 at such temperatures is negligibly small. We are not able to see any decay above 3 K for $H\perp c$. For this orientation, at 3 K, after an initial fall to a small value (stated above) till about 10 mins, the decay is insignificant. An apparent slow decay seen at 1.8 K for this orientation might signal the onset of glassiness in the close vicinity of about 2 K, which is difficult to confirm due to limited temperature window in our measurements. Even if glassiness is present below 2 K for this orientation, the findings are consistent with anisotropic nature of glassy magnetism, setting in well below $T_N$. We have fitted the curves for 1.8 and 8 K for $H//c$ to a stretched exponential of the type, $M_{IRM}(t) = M_{IRM}(0) + A \exp(-t/\tau)^{1-n}$. We find that the relaxation times, $\tau$, ~2000 s and 2700 s at 1.8 and 8 K are very large, mimicking the behavior of cluster spin-glasses, as discussed for polycrystals [21]. The value of n (= 0.5) is also consistent with this inference.

### 3.3. Ac susceptibility

Fig. 4 shows temperature dependence of ac $\chi$ for the two orientations, measured in the absence of any external magnetic field as well as in 5 kOe. For $H//c$, the real part ($\chi'$) increases with decreasing temperature in the paramagnetic state (Fig. 4(a)). As in low-field dc $\chi$ curves, there is a sharp upturn at $T_N$, followed by a peak at 17 K (e.g., for a measurement frequency of 1 Hz). It is clear from the figure 4a that there is a significant $\nu$-dependence of the values at lower temperatures, though the shift of the peak cannot be resolved clearly. A careful look at the curves in this figure reveals additional shoulders around 12.4 K and 5 K, as though the observed curve is a superposition of multiple curves. This suggests complexity of the spin-dynamics of the proposed glassiness. In the inset of Fig. 4(b), we show the Cole-Cole plot - the plot of $\chi'$ versus $\chi''$ - (for $\nu$ = 133 Hz). This plot is not a single semicircular arc, but it is made up of at least two arcs. This finding provides a support for more than one relaxation time characterizing the proposed glassiness. As a characteristic feature of spin-glasses, an application of a dc field of 5 kOe suppressed these curves, in particular the $\nu$-dependence, leaving a peak at the onset of long-range magnetic order. It is worth noting that the in-field and zero-field curves merge far above $T_N$ - near 45 K only - as in polycrystals, possibly due to the persistence of short-range magnetic order. Now, looking at the imaginary component ($\chi''$) (Fig. 4(b)), following a sharp upturn at $T_N$, there are peaks/shoulders at 6.4, 13.4 and 16 K in the curve for $\nu$= 1.3 Hz, which move towards higher temperatures with increasing $\nu$. Clearly, this finding supports the inferences made from the real part, with the temperature of the first peak, ~16 K (as one enters magnetically ordered state) in $\chi''$ at the lowest frequency representing the onset of glassiness. The observation of well-defined features in the imaginary part satisfies the necessary criterion for spin glass freezing [22]. In the presence of 5 kOe, the imaginary component completely vanishes, as expected for spin-glasses. Now turning to the ac $\chi$ behavior for $H\perp c$ (Figs. 4(c) and 4(d)), the curves are qualitatively different from that for $H//c$, with a pronounced reduction in absolute values (as in dc $\chi$). Notably, the curves for all the frequencies are found to overlap well above 2 K, though there is a mismatch close to 2 K, indicating possible onset of glassiness in the close vicinity of 2 K, as inferred from $M_{IRM}$. There is no well-defined peak in $\chi''$ down to 2 K. These features conclusively establish the absence of any glassiness of magnetism along basal plane well above 2 K.

### 4. Discussions

From the results on single crystals presented above, it is clear that the compound, $Tb_3Ru_4Al_{12}$, undergoes long-range antiferromagnetic ordering at 22 K. It appears that short-range magnetic correlations persist over a wide temperature range above $T_N$, as also inferred from the separation of zero-field and in-field $\chi'$ curves around 45 K for $H//c$. In addition, the bulk susceptibility data (both ac and dc) reveal the existence of additional features well below $T_N$. These findings on single crystals reinforces our earlier [20,21] assertion based on the work on the polycrystals that various collective states compete as a result of geometrical frustration in a kagome lattice – a point emphasized in a theoretical work on pyrochlores by Jaubert et al [23].

The observation to be noted is that the spin-glass anomalies are seen in ZFC-FC dc $\chi$, ac $\chi$, and $M_{IRM}$ at temperatures lower than $T_N$, for $H//c$. Though one is generally tempted to attribute such glassy features to possible crystallographic disorder, we believe that this may not be the case in this compound on the following grounds: (i) If disorder is responsible for glassy anomalies, one would naively expect that similar glassy features for $H \perp c$ also should have been seen at the same temperature as in $H//c$, in contrast to the observations. (ii) The absolute values of $\chi'$ (as well as of $\chi''$), for instance at the peak, are of comparable magnitude to that noted for polycrystals. As disorder and domain wall effects should be widely different for these two forms of the material, such comparable values for two forms of the compound offer is suggestive for the role of disorder-free effects, that is, geometrical frustration. (iii) The Gd analogue [19], even in the polycrystalline form, naturally with defects and domain walls, does not exhibit glassy features.

## 5. Conclusion

We have presented the results of ac and dc susceptibility measurements as a function of temperature on the single crystals of a metallic kagome lattice, $Tb_3Ru_4Al_{12}$. The bulk magnetization results reveal anisotropic spin-glass features below about 17 K and long-range antiferromagnetic order setting in at 22 K. Therefore, further studies on this compound would be worthwhile to explore whether this compound serves as a model system for the proposal of Chandra et al [5] mentioned at the introduction, and also to understand spin-glass features in geometrically frustrated metallic kagome lattices.

**Acknowledgement**
One of the authors (A.V.A) acknowledges the support of the Materials Growth and Measurement Laboratory, and grant 16-03593S of the Czech Science Foundation, for the single crystal growth.

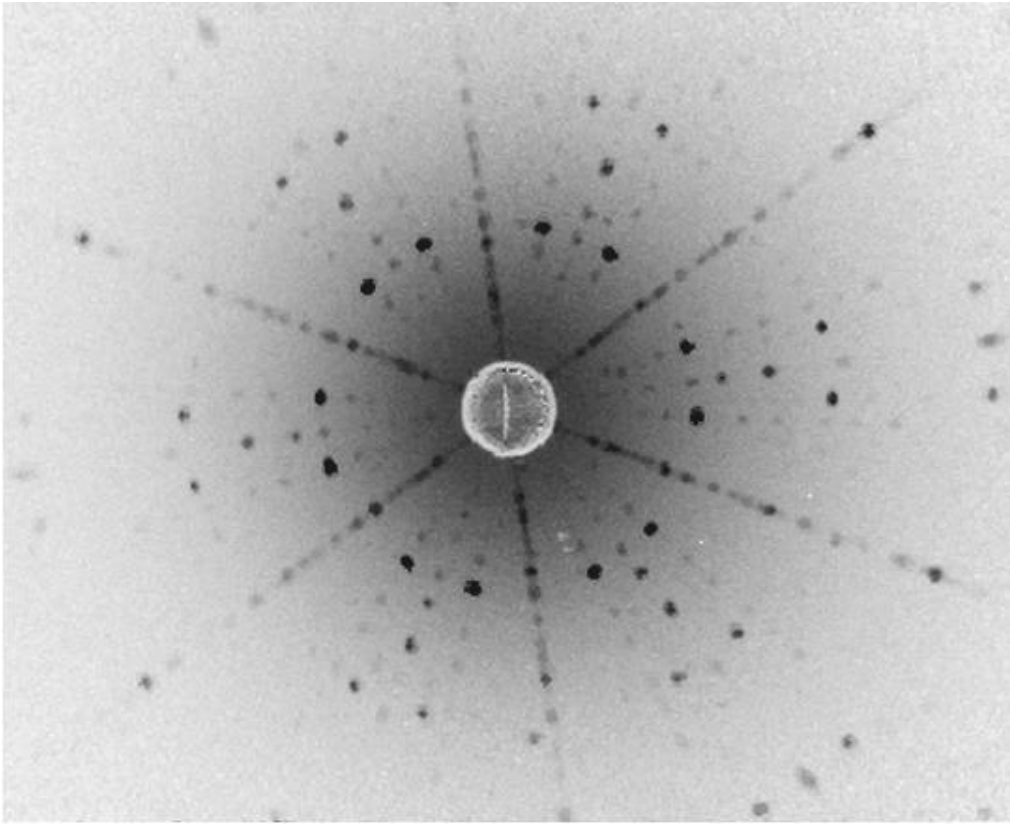

**Fig. 1**: Back-scattered Laue pattern along the c-axis of the $Tb_3Ru_4Al_{12}$ single crystal.

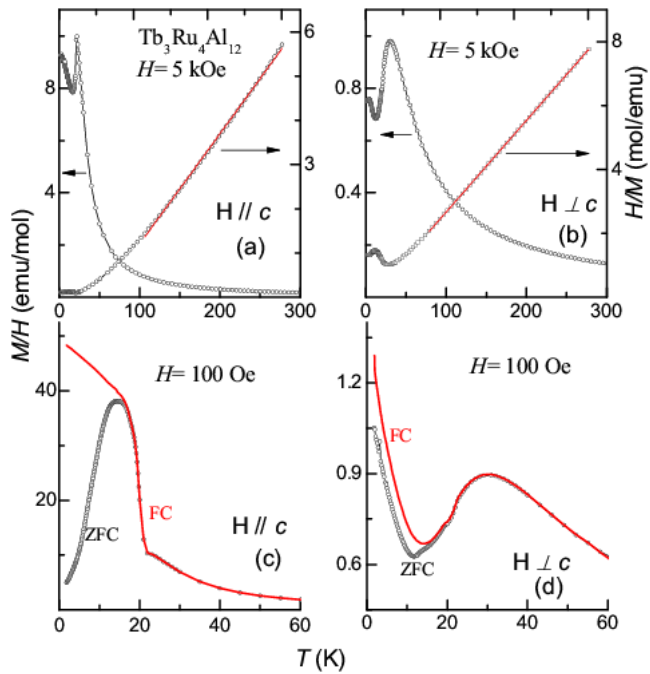

**Fig. 2:** Temperature dependence of magnetic susceptibility ($\chi = M/H$) and inverse magnetic susceptibility of single crystalline Tb$_3$Ru$_4$Al$_{12}$ for two orientations with respect to magnetic field in 5 kOe and 100 Oe. The line through the data points serve as guides to the eyes, except in inverse $\chi$ plots, in which case the line represents a Curie-Weiss fit in the high temperature range. While in (a) and (b), the curves belong to zero-field-cooled (ZFC) condition, in (c) and (d), the curves are shown for both ZFC and FC conditions.

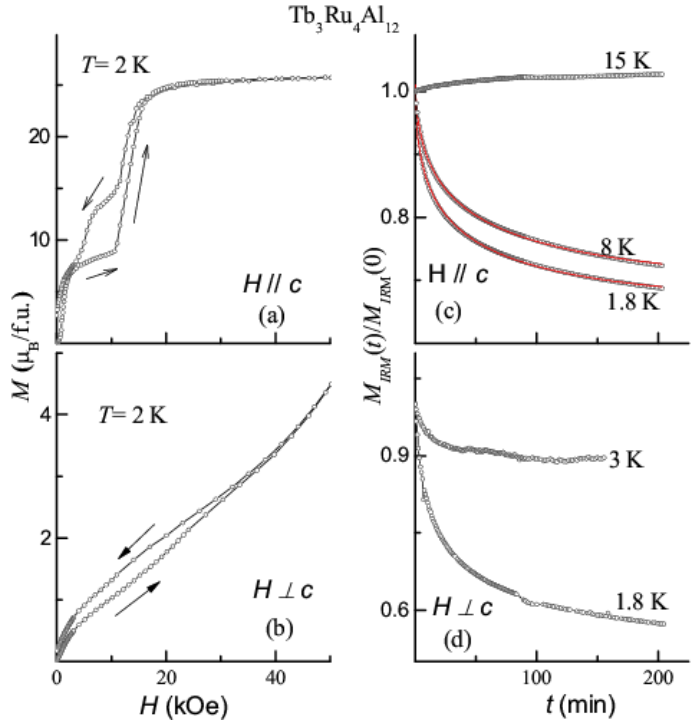

**Fig. 3:** Isothermal magnetization behavior of single crystalline Tb$_3$Ru$_4$Al$_{12}$ at 2 K for the orientations, (a) $H//c$ and (b) $H\perp c$. For the two orientations, isothermal remnant behavior as a function of time at different temperatures is shown in (c) and (d) respectively. The lines through the data points serve as guides to the eyes, except in (c) in which the lines are obtained by a fitting to a stretched exponential function, as described in the text.

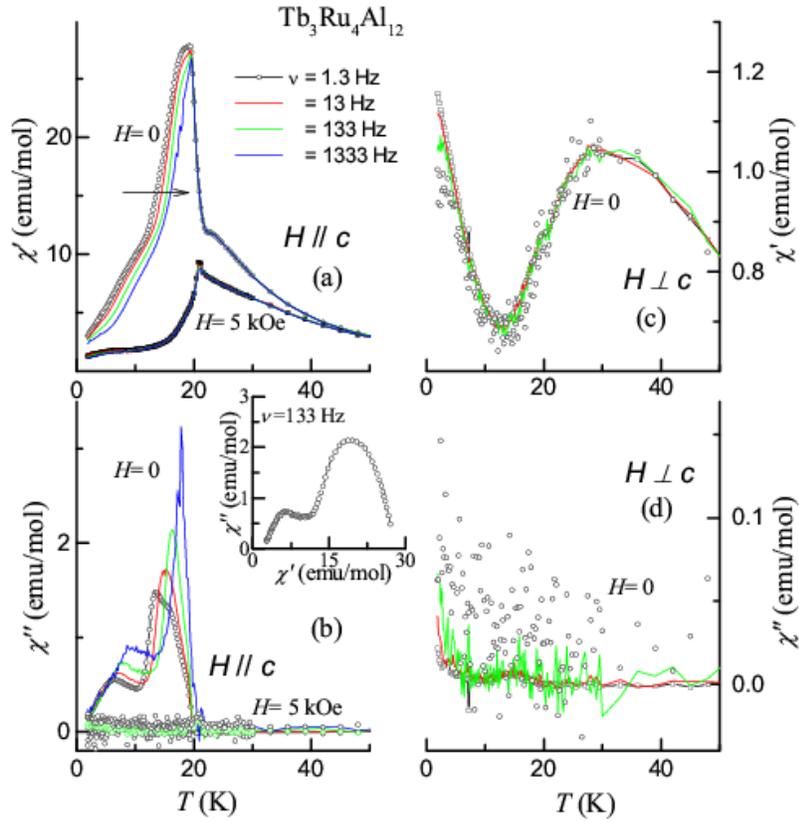

**Fig. 4:** Real ($\chi'$) and imaginary ($\chi''$) parts of ac susceptibility as a function of temperature for single crystalline Tb$_3$Ru$_4$Al$_{12}$ for two orientations $H_{ac}//c$ (left) and $H_{ac}\perp c$ (right), measured with frequencies ($\nu$) 1.3, 13, 133 and 1333 Hz. For the sake of clarity, the data points are shown for $\nu=$ 1.3 Hz only. The data obtained in dc H= 5 kOe in (b) and in the absence of dc magnetic field in (d) are featureless. Horizontal arrow in (a) show the direction in which the curves move with increasing $\nu$. Cole-Cole plot is shown in the inset of (b).